\documentclass[aps,prb,twocolumn,showpacs,floatfix,superscriptaddress]{revtex4-1}
\usepackage[dvips]{graphicx}
\usepackage{dcolumn}
\usepackage{amsmath}
\usepackage{amssymb}
\usepackage{multirow}
\usepackage[english]{babel}
\usepackage{color}
\usepackage{moreverb}
\usepackage{bm}
\newcommand{\g}{graphene}

\def\bra#1{\mathinner{\langle{#1}|}}
\def\ket#1{\mathinner{|{#1}\rangle}}

\newcommand{\etal}{\textit{et al.\/}}

\begin{document}
\title {Fe clusters (Fe$_n$, $n$=1-6) chemisorbed on vacancy defects in graphene: Stability, spin-dipole moment and magnetic anisotropy} 

\author{Soumyajyoti Haldar}
\email{Soumyajyoti.Haldar@physics.uu.se}
\affiliation {Division of Materials Theory, Department of Physics and Astronomy, Uppsala University, Box-516, SE 75120, Sweden}

 \author {Bhalchandra S. Pujari}
 \affiliation{Centre for Modeling and Simulation, University of Pune, Pune-411007, India}
 
 \author {Sumanta Bhandary}
\affiliation {Division of Materials Theory, Department of Physics and Astronomy, Uppsala University, Box-516, SE 75120, Sweden}
 
 \author {Fabrizio Cossu}
 \affiliation{Materials Science Division, Department of Physical Science and Engineering, KAUST, 23955-6900 Thuwal, KSA}

\author{Olle Eriksson}
\affiliation {Division of Materials Theory, Department of Physics and Astronomy, Uppsala University, Box-516, SE 75120, Sweden}
 
\author {Dilip G. Kanhere}
 \affiliation{Centre for Modeling and Simulation, University of Pune, Pune-411007, India}

\author {Biplab Sanyal}
\email[Corresponding Author: ]{Biplab.Sanyal@physics.uu.se}
\affiliation{Division of Materials Theory, Department of Physics and Astronomy, Uppsala University, Box-516, SE 75120, Sweden}

\begin{abstract} 
    
In this work, we have studied the chemical and magnetic interactions of Fe$_n$;
$n$=1-6 clusters with vacancy defects (monovacancy to correlated vacancies with six missing C atoms) in a graphene sheet by ab-initio density functional calculations combined with Hubbard U corrections for correlated Fe-d electrons. It is found that the vacancy formation energies are lowered in the presence of Fe, indicating an easier destruction of the graphene sheet. Due to strong chemical interactions between Fe clusters and vacancies, a complex distribution of magnetic moments appear on the distorted Fe clusters which results in reduced averaged magnetic moments compared to the free clusters.  In addition to that, we have calculated spin-dipole moments and magnetic anisotropy energies. The calculated  spin-dipole moments arising from anisotropic spin density distributions, vary between positive and negative values, yielding increased or decreased effective moments. Depending on the cluster geometry, the easy axis of magnetization of the Fe clusters shows in-plane or out-of-plane behavior.

\end{abstract}

\maketitle

\section{Introduction}

Manipulation of the properties of graphene \cite{novoselov} by controlled defect insertion is one of the focused research topics related to graphene.  Removal of carbon atoms by ion-irradiation gives rise to single or correlated vacancies and hence, the transport properties have been shown to be highly modified. \cite{jafri} At the same time, the route of chemical functionalization of graphene has been quite appealing too in the endeavor of opening band gaps, e.g., in the form of hydrogenated graphene (graphane) \cite{elias01302009} or graphene sheets chemically modified by fluorine atoms. \cite{Nair:2010eh, Robinson:2010ki} Moreover, graphene offers the unique possibility for foreign elements, e.g., adatoms, molecules, clusters etc. to adhere to its extended surface and delocalized $\pi$-electron band. The presence of defects in the graphene sheet facilitates the adsorption and modification of the properties of graphene. The nature and types of defects in {\g} have been discussed by Castro Neto {\it et. al.} in their extensive review. \cite{netormp} Among a number of defects and disorders seen in {\g}, ripples or topological defects are intrinsic while cracks, vacancies, charged impurities, atomic adsorption etc. are extrinsic.  In particular {\g} is prone to the formation of vacancy defects. Divacancy defects are quite probable to form as strong reactive centers. \cite{biplabprb09, nieminen} It is shown theoretically \cite{lust} how different defect structures can be engineered in graphene.  It is known that such defects can affect  the electronic structure and hence transport properties of graphene \cite{coleman, jafri, carva}.

A considerable attention has been paid to explore graphene in modifying the magnetic properties of adatoms and clusters and vice-versa. Recent theoretical studies show that it is possible to attach metal atoms in defected graphene much more firmly than to pristine graphene~\cite{k-dadu,k-dadu-prl,kim}. Electronic and magnetic properties of transition metal (TM), e.g., Fe, Co, Ni clusters on monolayer graphene have been studied using X-ray Magnetic Circular Dichroism (XMCD)~\cite{tim-prl}. Also it has been predicted that a Ir dimer adsorbed in single-vacancy graphene has a very high magnetic anisotropy energy (MAE) ~\cite{gmae}. It has also been reported that magnetic properties of graphene can be modified through combined TM and nitrogen decoration of vacancies. Dynamics of single Fe adatoms on graphene vacancies has been studied using focused electron beam irradiation. ~\cite{1fe-dyna} From the perspective of a non-covalent interaction between graphene and magnetic species, recent theoretical studies \cite{sumantaprl, sumantansr} have suggested that the magnetic properties, such as spin moments, spin dipole moments and magnetic anisotropy energies can be modified by the defects present in graphene. 

One motivation to introduce magnetism in graphene is to explore the possibility of having spin-dependent transport. One may try to adsorb magnetic adatoms on graphene and study the range of spin polarization in the host lattice arising from the exchange coupling between the adatoms. It is furthermore possible to make use of the defects for trapping the adatoms. It is known that the chemisorption energies at vacancy sites are very high \cite{biplabprb09}. So it is possible to trap magnetic adatoms or clusters at various defect sites with the ambition to have an effective spin polarization. In an interesting theoretical work based on density functional theory (DFT), Krasheninnikov {\it et. al.} \cite{nieminen} have studied the magnetism of different transition metal adatoms on single and divacancy centers. They have shown that for most metals, the bonding is strong and the metal-vacancy complexes exhibit interesting magnetic behavior. In a work based on DFT, Wang {\it et al} \cite{wang} have studied the interaction between a single adatom and graphene containing a Stone-Wales defect. They have observed a reduction in local magnetic moment on the Fe atom and a substantial modulation of electronic states near the Fermi level. Instead of a single adatom, a flux of adatoms may generate various sizes of magnetic nanoclusters trapped at defect sites and therefore, they are a topic of interest. How robust the formation of cluster on defected sites is also a topic of interest and we have done Born-Oppenheimer molecular dynamics (BOMD) to analyze that. 

Regarding spintronic applications, magnetic anisotropy energies (MAE) play an important role. One of the most important points is to achieve out-of-plane magnetization (e.g. for 2D system like graphene, it is perpendicular to the surface of the sheet) for increasing storage density. Finite systems, such as magnetic clusters may act as potential candidates in this regard. Magnetic adatoms on graphene have been studied recently. In a recent report, strong magnetic anisotropy of chemically bound Co dimers in a graphene sheet has been discussed.\cite{kandpal}. DFT calculations \cite{porter} also revealed an in-plane easy axis of magnetization for an isolated Fe adatom on graphene. However, in this study, strong Coulomb correlations have been neglected. Motivated by these studies, we have investigated MAEs of Fe clusters of various sizes and their effects of stronger interactions with vacancy sites with the inclusion of strong Coulomb interactions. Also, it has been shown recently \cite{stefan} that the magnetic anisotropy of adatoms on graphene can be tuned by an electric field. In fact, it was shown that the easy axis of magnetization can be modified from in-plane to an out-of-plane direction in this route, which could be appealing for technological developments. Hence, magnetic anisotropy and its modification by external fields is an important topic of research.

There is another aspect that needs attention; namely, the spin-dipole contribution, which arises from the anisotropic spin densities. It is usually neglected for bulk systems with high symmetry. Recently, it has been shown that for low-dimensional systems, e.g., clusters \cite{hubert} and organometallic molecules with magnetic centers, \cite{gambardella, sumantaprl,sumantaprb} this contribution, commonly denoted as $<T_z>$ has a significant role. More interestingly, the spin-dipole contribution may have an opposite sign to that of spin moment, thereby reducing the effective moment ($m_{eff}=m_{s}+7<T_z>$), which is measured in XMCD experiments. \cite{heikeprb} Therefore, the Fe clusters are expected to have interesting variations in spin-dipole contributions due to their anisotropic shapes and spin densities. 

The present work is a DFT study of the adsorption of small clusters of Fe (Fe$_n$ ; $n = 1 - 6$) on vacancy sites, with a varying size of the vacancy center by removing $n$ C atoms ($n = 1 - 6$). The plan of the paper is as follows: In the next section we will present computational details followed by the results in section \ref{sec:results}. We will discuss briefly about the diffusion of number of Fe adatoms on a defected 2D graphene sheet using BOMD in section \ref{subsec:md}. Then in section \ref{subsec:vacd}, we will discuss different vacancy defects that we have considered in the present study and then we will proceed to the discussion of Fe$_n$ clusters on $n$ vacancies in section \ref{subsec:fe-gr}. Spin-dipole moments and orbital moments will be presented in section \ref{sub2sec:spin-dipole} and \ref{sub2sec:mae} respectively. The conclusions are given in section \ref{sec:conclude}

\section{\label{sec:compdetail}Computational details}
All the calculations have been performed on a monolayer graphene supercell. The varying sizes of vacancy defects are created by removing a cluster of C atoms ranging from 1 to 6 (designated by V$_{1}$ to V$_{6}$). The C atoms are removed from hexagonal ring in such a way that the vacancies are always on adjacent sites. Then the Fe$_n$ cluster is placed on a $n$-vacancy. The supercell is generated by repeating the primitive cell by six times in $a$ and $b$ direction. Such large unit cell is required to avoid the interaction between the clusters. The effect of vertical interaction is avoided by taking a vacuum of 16 {\AA}.  All the calculations have been performed using plane-wave based density functional code \textsc{vasp}. \cite{vasp}  The generalized gradient approximation as given by Perdew, Burke and Ernzerhof ~\cite{PBE,PBEerr} has been used  for the exchange-correlation potential.  The structures were optimized using the conjugate gradient method with forces calculated from the Hellman-Feynman theorem. The energy and the Hellman-Feynman force thresholds are kept at 10$^{-5}$ eV and 10$^{-2}$ eV/{\AA} respectively. For magnetic anisotropy energy (MAE) calculation, we have used  10$^{-6}$ eV energy cutoff. For geometry optimization, a 5 $\times$ 5 $\times$ 1 Monkhorst-Pack $k$-grid is used. Total energies and electronic structures are calculated with the optimized structures on 11 $\times$ 11 $\times$ 1  Monkhorst-Pack $k$-grid. Diffusion of Fe adatoms on a defected 2D graphene sheet at 300 K temperature was studied using Born-Oppenheimer molecular dynamics (BOMD) simulations. The temperature was adjusted via a Nos\'{e} thermostat~\cite{nose1,nose2,nose3}.

The vacancy formation energy $E^{vac}$ of a n-vacancy graphene sheet with reference to pristine graphene sheet is defined as follows
\begin{equation}
\label{eq:vac_for}
 E^{vac} = E_{\rm V_n} - \frac{\rm N-n}{\rm N}E_{\rm Gr}
\end{equation}
where $E_{\rm V_n}$ is the total energy of n-vacancy graphene sheet, $E_{\rm Gr}$ is the total energy of pristine graphene sheet and N is the total number of C atoms in the pristine graphene. 

The vacancy formation energy $E^{vac}_{Fe}$ of n-vacancy graphene sheet in presence of Fe$_n$ cluster is defined as follows. 
\begin{equation}
\label{eq:vac_fe}
 E^{vac}_{Fe} = E_{\rm V_n + Fe_n} - \big [ E_{\rm Gr + Fe_n} - {\rm n}\frac{E_{\rm Gr}}{\rm N}\big]
\end{equation}
$E_{\rm V_n + Fe_n}$ is the total energy of the Fe$_n$ cluster adsorbed on the n-vacancy graphene sheet, $E_{\rm Gr + Fe_n}$ is the total energy of the Fe$_n$ cluster adsorbed on the pristine graphene sheet.

The adsorption energy $E^{ad}$ of Fe$_n$ cluster on the n-vacancy graphene sheet is defined as
\begin{equation}
\label{eq:add_fe}
E^{ad} = [E({\rm Fe_n}) + E({\rm sheet})] - E({\rm Fe_n+sheet}),
\end{equation}
where $E({\rm Fe_n+sheet})$ is the total energy of the Fe$_n$ cluster adsorbed
on the n-vacancy graphene sheet, $E({\rm sheet})$ is the total energy of the n-vacancy graphene sheet, and $E({\rm Fe_n})$ is the total energy of an isolated Fe$_n$ cluster kept in a big box.

For Fe nanostructures, electron correlation effects are expected to be important due to narrow band widths. In order to check what effects such electron correlations play in the electronic properties of the Fe$_n$ clusters on the n-vacancy graphene sheet, we have used the PBE+U method~\cite{ldau}, where the Coulomb parameter U is added in the Hubbard formalism. For these calculations we used U=4 eV and the intra-atomic exchange parameter J=1 eV, which are typical values for $3d$ transition metals.~\cite{ldau,uj1}

\section{\label{sec:results}Results and discussion}
\subsection{\label{subsec:md} Diffusion of Fe adatoms on defected graphene}
It is important to investigate how a number of Fe adatoms placed on a defected graphene diffuse, especially whether the Fe adatoms remain isolated or form a cluster to finally get trapped at the defect sites. In order to answer this, we have performed limited real time Born-Oppenheimer molecular dynamics at constant temperature $T=300$ K, a temperature expected in most practical applications. 
\begin{figure*}[ht]
  \begin{center}
    \includegraphics[scale=0.7]{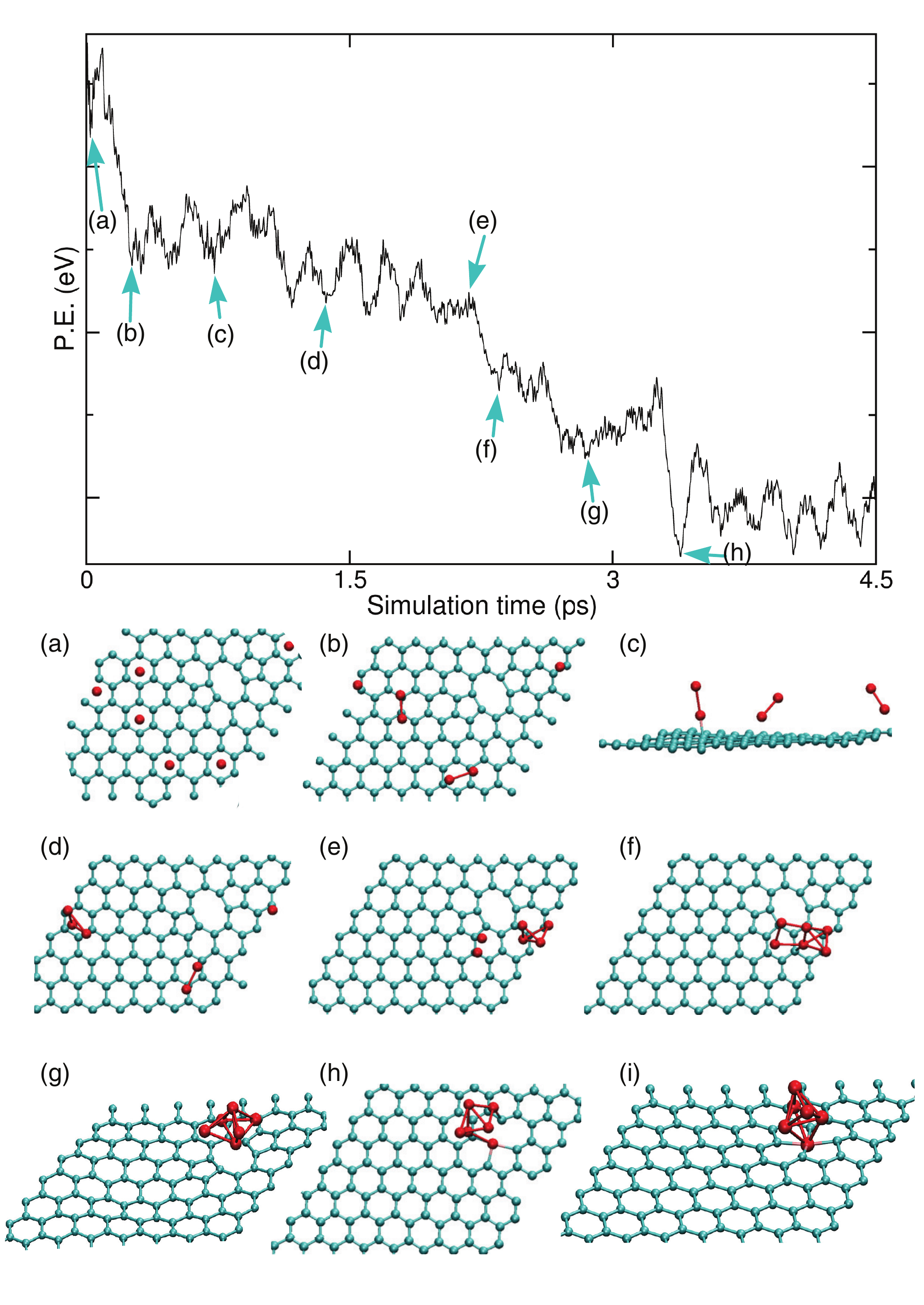}
  \end{center}
  \caption{(Color online) Potential energy (P.E.) landscape as a function of simulation time. Figs. (a)-(h) show the snapshots at different simulation times. These snapshots are also marked by arrow in the P.E. graph. Fig. (i) shows the optimized geometry after the completion of MD simulation for 30 ps. Cyan and red balls indicate C and Fe atoms respectively. See text in~\ref{subsec:md} for more details.}
  \label{fig:md1}
\end{figure*}
We have placed six Fe adatoms at six different hexagonal sites on the sheet and also removed two adjacent C atoms to create a divacancy. Then we performed finite temperature BOMD up to 30 ps simulation time. Figure \ref{fig:md1} shows the potential energy landscapes and corresponding geometries (marked by arrows in figure) at different simulation times. From the analysis of trajectories and movies of the motion of Fe atoms, a number of interesting observation emerged. Firstly, at the very beginning of the simulation, the C atoms near the divacancy center come closer and heal the divacancy to form a 5-8-5 defect (fig. \ref{fig:md1}(a)). Then the Fe adatoms from the adjacent hexagonal sites come together and form dimers (fig. \ref{fig:md1}(b)-\ref{fig:md1}(d)) in about $< 0.7$ ps. These dimers come closer to form a cluster (fig. \ref{fig:md1}(e)-\ref{fig:md1}(g)). Finally, in the presence of the Fe cluster near to the defect, one Fe atom from the cluster forms bond with the C atoms at the defect site and hence breaks the C-C bonds of the pentagonal rings of the 5-8-5 defect (fig. \ref{fig:md1}(h)). Once the cluster gets trapped at the defected site, it remains trapped and only shows thermal oscillations. The cluster remains stuck at the defect site and thermally oscillates during the remaining simulation time. Frequent breaking and forming of bonds within the Fe cluster and between Fe and C atoms can be seen during this time. Fig.~\ref{fig:md1}(i), the lowest energy structure obtained from the geometry optimization after the MD simulation, shows the formation of bonds within Fe cluster and between Fe and C atoms. Thus our simulation shows that by and large Fe atoms placed on defected graphene accumulate most preferably via dimerization to form clusters that get trapped at the defected site. Our simulation indicates that formation of cluster at defected site is fairly robust. Therefore it becomes more important to investigate in depth the interaction of Fe clusters with the defected graphene.

\subsection{\label{subsec:vacd}Vacancy defects in graphene}
\begin{figure*}[ht]
\includegraphics[scale=0.5]{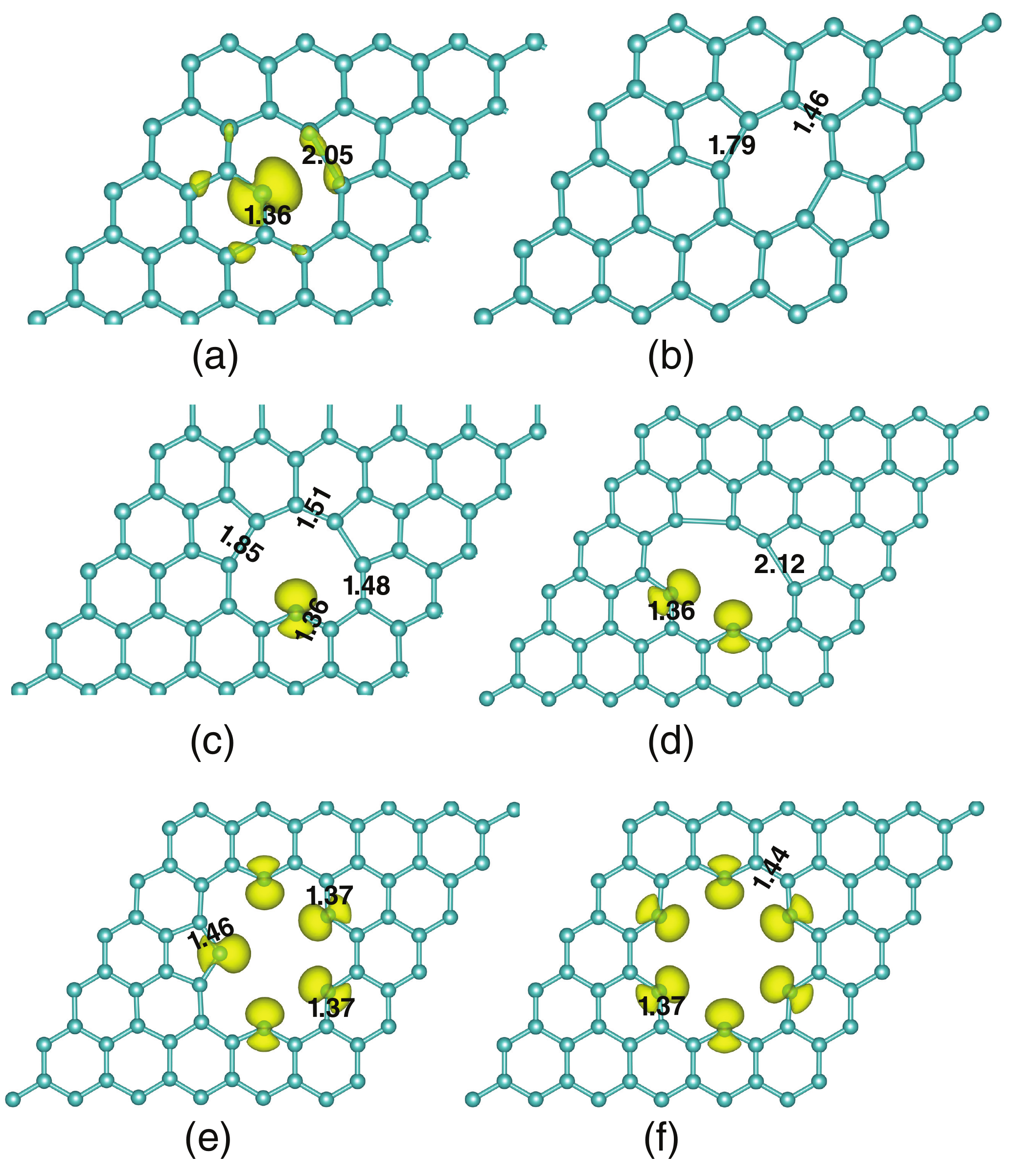}
\caption{\label{fig:vacancy} (Color online) Geometries and spin densities of vacancy defects in graphene.  Figure (a) to (f) represents the vacancy V$_1$ to V$_6$ respectively. Yellow isosurface denotes the spin density of the system. The numbers on each figure denotes the bond length in {\AA}.}
\end{figure*}
We begin our discussion by presenting the results of the vacancy defects V$_{1}$ to V$_{6}$ in graphene. The vacancies are created by removing C atoms from the same hexagonal ring. Fig. \ref{fig:vacancy} shows geometries and spin densities of these six different vacancy defects. 
\begin{table}[ht]
\caption{Vacancy formation energies $E^{vac}$ (in eV), total magnetic moments and vacancy formation energies $E^{vac}_{Fe}$ in presence of Fe$_n$ cluster.\\~\\~\\}
\label{tab-vac-energy}
\centering
\begin{tabular}{||r|c|c||c||}
\hline
 n &  $E^{vac}/\rm n$  & $\mu_{B}$ & $E^{vac}_{Fe}/\rm n$  \\
  &  (total)  &  total & (total)   \\
\hline
      1  & 6.92 &   1.48   & 2.73  \\
      2  & 3.54 &   0.00   & 1.49  \\
      3  & 3.44 &   1.00   & 1.19  \\
      4  & 3.38 &   2.00   & 0.87  \\
      5  & 3.17 &   5.72   & 0.69  \\
      6  & 2.73 &   6.00   & 0.65  \\
\hline
\end{tabular}
\end{table}
Vacancy formation energies (pristine graphene and Fe$_n$ on pristine graphene as reference) and total magnetic moments are tabulated in Table \ref{tab-vac-energy}.  A single vacancy in pristine graphene yields a total magnetic moment of 1.48 $\mu_B$ which largely comes from the dangling bond (as seen from the spin densities in fig. \ref{fig:vacancy}(a)) of the C atom. There are local arrangements in C-C bond lengths to form a pentagonal ring. In V$_2$ the divacancy defect heals to form a 5-8-5 defect with two pentagons and one octagon. All the dangling bonds saturate to give zero magnetic moment. In both V$_3$ and V$_4$, there are two pentagonal rings formed due to local rearrangements and the magnetic moment (1 $\mu_B$ and 2 $\mu_B$ respectively) can be seen on C atoms containing the dangling bonds (fig. \ref{fig:vacancy}(c)-\ref{fig:vacancy}(d)). In spite of a pentagon formation in V$_5$, there are five C atoms with dangling bonds giving rise to a total 5.72 $\mu_B$ magnetic moment. There is also no saturation of dangling bonds in V$_6$ and thus it gives a total magnetic moment of 6 $\mu_B$. The analysis of spin-densities revealed that the  dangling bonds are formed from the in-plane $p$-orbitals of the C atoms. Therefore vacancy defects in graphene create varieties of locally distorted structures with localized magnetic moments. However, a general trend from Fig. \ref{fig:vacancy} is that the moment distribution gets more homogeneously distributed over the edge of the defect, as the size of the defect grows.

A few interesting observations can be made from Table \ref{tab-vac-energy} regarding the vacancy formation energies. The formation energies of the vacancies are gradually reduced with an increase in the number of vacancies correlated to each other indicating that already formed vacancies help to create more vacancies to make the hole bigger and bigger. However around V$_6$ it saturates and the vacancy formation energy (per vacancy) saturates at around 0.65 eV. Another important observation is that Fe clusters promote creation of vacancies as the formation energies are drastically reduced in presence of Fe.  This is consistent with the observation of Ref.~\onlinecite{boukhvalov:09}, where the Fe atoms were shown to destroy the graphene sheet by creating vacancies.

\subsection{\label{subsec:fe-gr}Fe$_n$ clusters on defected graphene}

\subsubsection{\label{sub2sec:fe-gr-geom}Analysis of geometries}
\begin{figure*}[ht]
\includegraphics[scale=0.5]{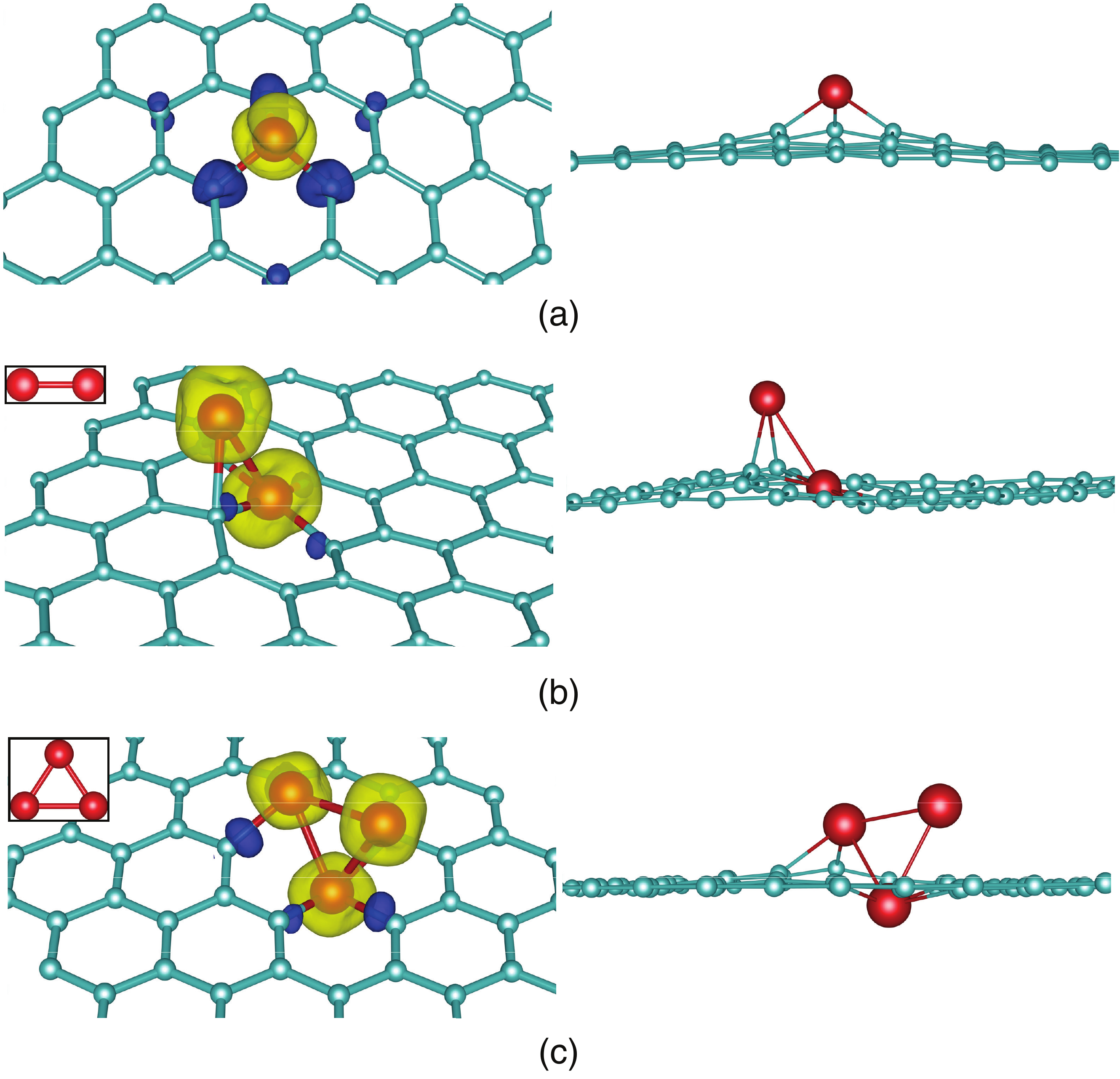}
\caption{\label{fig:fen-vn-1} (Color online) Geometries and spin density of Fe$_{n}$ clusters on V$_{n}$ vacancies (n=1-3) are shown together. Cyan and red balls indicate C and Fe atoms respectively. Yellow and blue isosurface denotes positive and negative spin densities. }
\end{figure*}
\begin{figure*}[ht]
\includegraphics[scale=0.5]{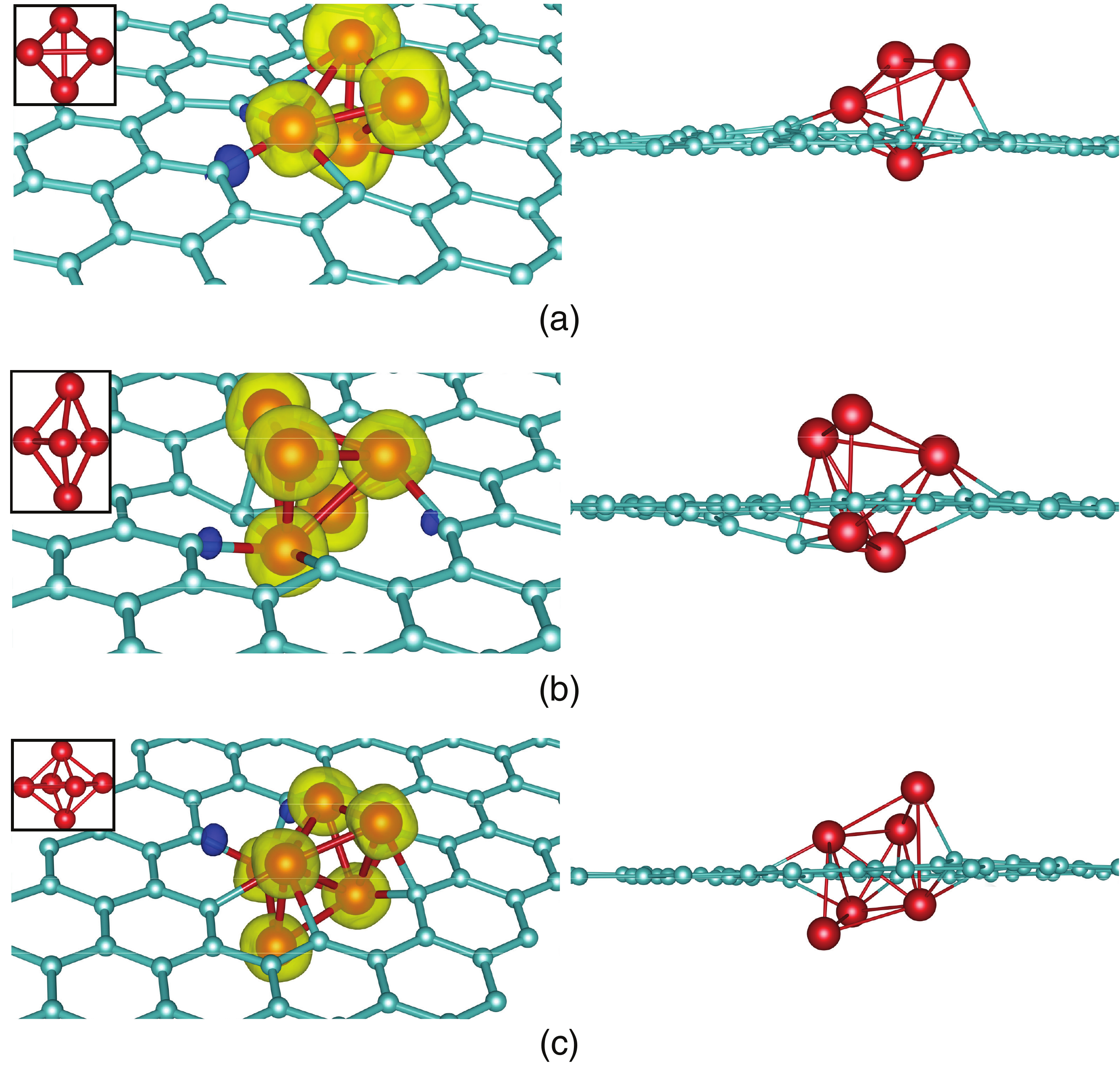}
\caption{\label{fig:fen-vn-2} (Color online) Geometries and spin densities of Fe$_{n}$ clusters on V$_{n}$ vacancies (n=4-6) are shown together. Cyan and red balls indicate C and Fe atoms respectively. Yellow and blue isosurface denotes positive and negative spin densities. }
\end{figure*}
Now we present the ground state geometries of Fe$_n$ clusters adsorbed on V$_n$. In Fig. \ref{fig:fen-vn-1} and \ref{fig:fen-vn-2} we have shown the geometries and spin densities for the systems with $n=1-3$ and $n=4-6$ respectively. The left and right column show top and side views respectively.  For the sake of comparison, the geometries of the free Fe clusters are also shown in the insets of the figures at the left column. As seen from Fig. \ref{fig:fen-vn-1}(a), the optimized geometry for Fe$_1$ on V$_1$ shows that the minimum energy position for Fe is at the center of the vacancy. The average Fe-C bond length is $\sim$ 1.77 {\AA} for this case and the height of the Fe adatom from the graphene plane is $\sim$ 1.19 {\AA}. For the Fe$_2$ on V$_2$, the two Fe atoms forms a dimer of length 2.20 {\AA} which is about 10\% enhanced from the bond length found in a free cluster (2.02 {\AA}). However the dimer geometry is peculiar in the sense that one Fe atom resides in the plane of graphene while the other Fe atoms is out of the plane (see figure \ref{fig:fen-vn-1}(b)). The four Fe-C bonds are identical with bond lengths of 1.97 {\AA}. It is interesting to note that due to the presence of the Fe atom in the graphene plane, the underlying vacancy does not undergo any significant structural rearrangement.  This is in contrast with the vacancy in absence of Fe atoms where the C atoms move close to each other to form in-plane $\sigma$ bonds, resulting in a 5-8-5 structure \cite{coleman}. The optimized geometry for trimer shows (see fig. \ref{fig:fen-vn-1}(c)) that the trimer plane is aligned perpendicular to the graphene plane. As the size of vacancy increases, one Fe adatom goes slightly below the graphene plane. The average Fe-Fe bond length is $\sim$ 2.3 {\AA} and the average Fe-C bond length remains the same as mentioned before.

As the size of vacancy and the number of Fe atoms grow beyond three, the most notable change in the geometry of Fe clusters occurs with the movement of one of the Fe atoms below the plane of graphene. Moreover, the geometries of the Fe clusters transform into three dimensional ones. As the size of the vacancy increases, the first system featuring 3D Fe clusters, Fe$_4$, has also a larger vacancy to fill. Consequently, to maximize bonding with the C atoms, one of the Fe atoms of Fe$_4$ goes below the graphene plane (fig. \ref{fig:fen-vn-2}(a)). A similar behavior can also be seen for $n=5 \text{ and } 6$. For $n=5$ and $n=6$, two and three Fe atoms go below the graphene plane respectively (as seen in fig. \ref{fig:fen-vn-2}(b)-\ref{fig:fen-vn-2}(c))

Our analysis of geometry indicates that in general, the underlying vacancy does not undergo any significant structural rearrangement in presence of Fe clusters. Thus even a single Fe adatom can be used to maintain the structure of underlying lattice despite the presence of the vacancy. The geometries of Fe clusters on graphene are remarkably different than those studied in free space~\cite{yu}.  As mentioned earlier the dimer bond length is slightly reduced in presence of a divacancy in  graphene. The trimer in free space is reported to be a isosceles triangle \cite{yu,gustev}, while the trimer on a vacancy is a distorted isosceles triangle with the bond lengths differing substantially from that in free space. Fe$_4$ also forms a prism in free space with the bond lengths ranging from 2.22-2.41 {\AA}. However in our case the bond length varies substantially from 2.2 {\AA} to 2.63 {\AA}. The change in the trigonal bi pyramid of isolated Fe$_5$ is seen in the vertical four-atom plane where the bond lengths are increased with respect to those in free space.  Fe$_6$ undergoes a substantial change from octahedron to a more complex structure seen in fig. \ref{fig:fen-vn-2}(c).

\subsubsection{\label{sub2sec:fe-gr-energy}Energetics and Magnetic structure}
Next, we analyze the energetics and magnetic structures. In Table \ref{tab-ads-energies}, the adsorption energies for the Fe$_n$ clusters in a $n$-vacancy graphene sheet and their magnetic moments are reported, together with the magnetic moments of free Fe$_n$ clusters. As seen from the table \ref{tab-ads-energies}, the magnetic moments of Fe clusters are slightly reduced by their adsorption on the $n$-vacancies except for Fe$_6$ on V$_6$ and Fe$_1$ on V$_1$ where the reduction is huge. However, as seen from the spin density figures (fig. \ref{fig:fen-vn-1} and fig. \ref{fig:fen-vn-2}) the magnetic character of the C atoms near to the vacancy is significantly affected. The magnetization of the whole system is still basically due to the Fe adatoms. Except for the case with one Fe adatom on a single vacancy, the average Fe magnetic moments for the adsorbed clusters are not so different from the moments in free state. The drastic reduction of magnetic moment for the Fe atom adsorbed at the monovacancy site is due to a strong hybridization between the Fe-d and C-p orbitals at the defect site. The analysis of local magnetic moment on an individual Fe atom clearly shows that the spin moments retain high values (shown later).
\begin{table}[ht]
\caption{Adsorption energies $E^{ad}_{V_n}$ (in eV) for Fe$_n$ cluster adsorbed in $n$-vacancy graphene sheet, together with their magnetic moments (in $\mu_{B}$). The calculations are done within PBE+U formalism. The middle column shows the total moment of an Fe cluster on graphene with defects, while the right hand column shows that calculated moment for free Fe clusters.}
\label{tab-ads-energies}
\centering
\begin{tabular}{|r||c|c||c|c||c|}
\hline
  &  $E^{ad}_{V_n}/\rm n$  & $\mu_{B} / \rm n$ &  $\mu^{Fe_n}_{B} / \rm n$ \\
  &     &    &   (Free) \\
\hline
      1  &  5.36   &   0.32  &     4.0 \\
      2  &  2.80   &   3.00  &     4.0 \\
      3  &  2.60   &   3.32  &     4.0 \\
      4  &  2.97   &  3.00  &   3.5 \\
      5  &  2.71   &  3.20  &    3.6 \\
      6  &  2.34   &  3.33  &    3.33 \\
\hline
\end{tabular}
\end{table}

It can be noted that magnetic moments of the C atoms for the defected graphene has reduced significantly after the adsorption of Fe clusters. A careful analysis reveals that the moments in the edge C atoms around the vacancy sites appear due to partially filled in-plane p orbitals. These orbitals are hybridized with the d-orbitals of adsorbed Fe atoms to pacify the dangling bonds and hence, destroy the local moments at those edge C atoms.
\begin{figure}[ht]
\begin{center}
\includegraphics[scale=0.35]{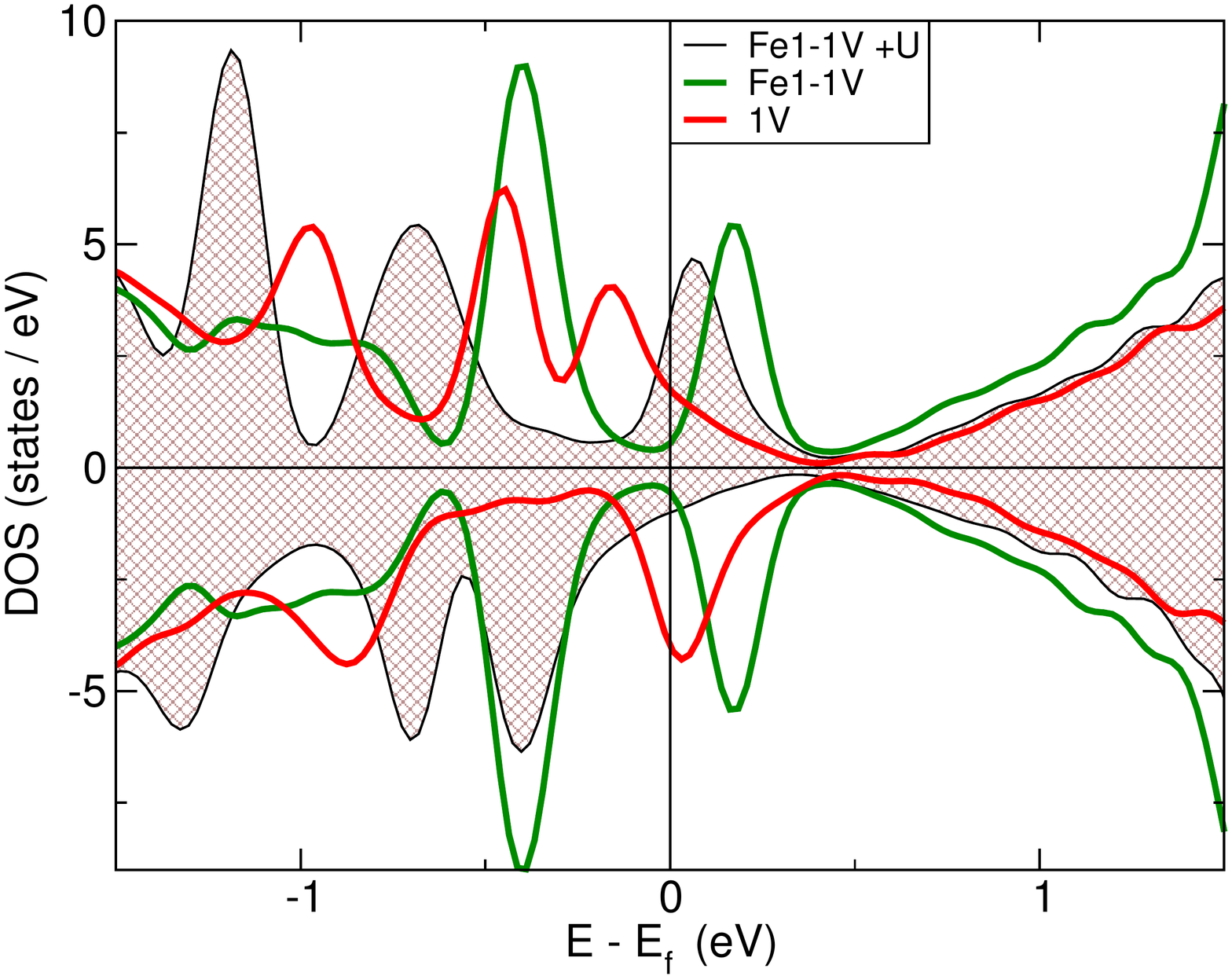}\\
{\bf (a)}\\
\includegraphics[scale=0.35]{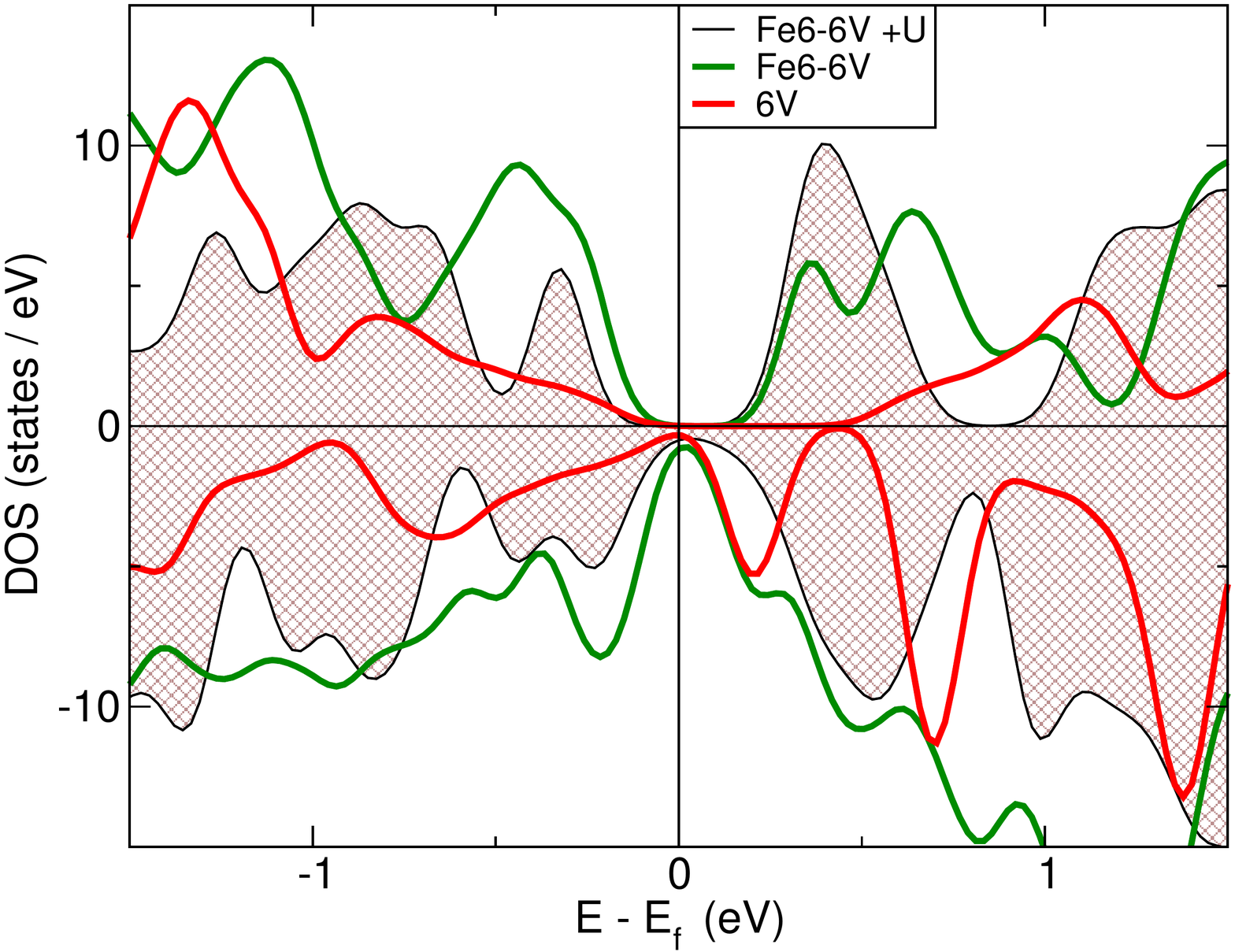}\\
{\bf (b)}
\end{center}
\caption{(Color online) Comparison of total DOS for (a) Fe$_1$ on V$_1$ and (b) Fe$_6$ on V$_6$. We have plotted DOS for pure vacancy (red lines) and the Fe cluster adsorbed in vacancy without U (green lines) and with U (shaded region).}
\label{fig:totaldos}
\end{figure}
\begin{figure}[ht]
\begin{center}
\includegraphics[scale=0.35]{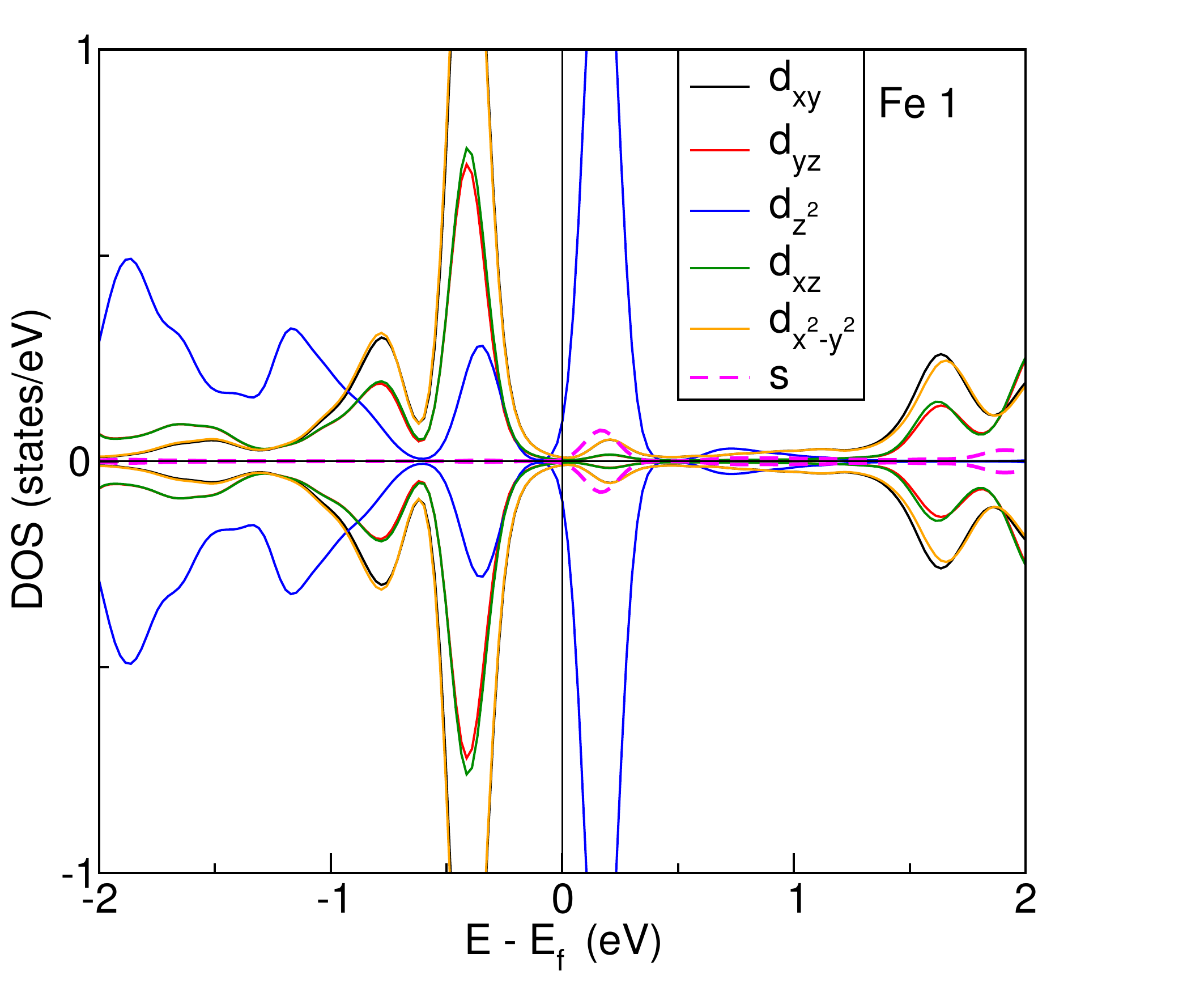}\\
{\bf (a) PBE}\\~\\~\\
\includegraphics[scale=0.35]{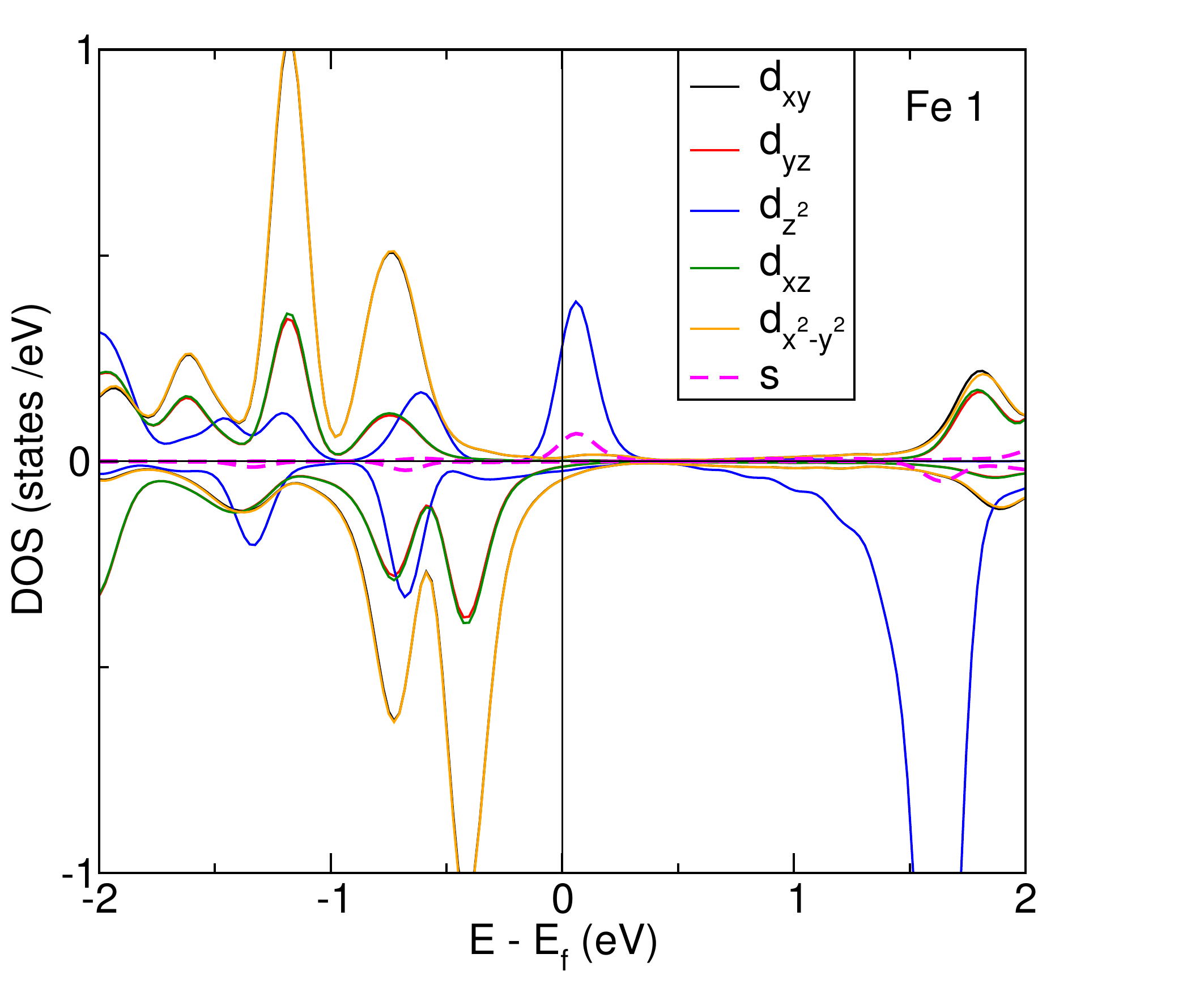}\\
{\bf (b) PBE+U}\\
\end{center}
\caption{(Color online) $m_l$ decomposed DOS for Fe atom for the system Fe$_1$ in 1-vacancy. Fig. (a) and (b) are within PBE and PBE+U formalism respectively.}
\label{fig:fe-ml-dos-1}
\end{figure}

The effect of Fe adsorption in the electronic structure is shown in Fig.~\ref{fig:totaldos}. Fig.~\ref{fig:totaldos}(a) and fig.~\ref{fig:totaldos}(b) show the comparison of density of states (DOS) for Fe$_1$ on V$_1$ and Fe$_6$ on V$_6$ respectively. The red and green lines represent DOS for pure vacancy and Fe adsorbed system within PBE formalism respectively. The shaded regions represent Fe adsorbed system within PBE+U formalism. Naturally in the presence of Fe clusters, the DOS of the graphene lattice undergoes a substantial change. Particularly interesting is the fact that the contributions from the Fe atoms occur directly at the Fermi level. Fig. \ref{fig:totaldos}(a) and fig. \ref{fig:totaldos}(b) show comparison of total DOS for defected graphene and Fe$_n$ cluster adsorbed on V$_n$ (both within PBE and PBE+U) for $n=1$ and $n=6$ respectively. Without U, the Fe-d states are too delocalized to form a local magnetic moment whereas with the inclusion of U, there is an enhancement in the exchange splitting and a small local moment is stabilized. In the bottom panel of Fig.~\ref{fig:totaldos}, one observes a peculiar DOS for the hexavacancy with a broad gap in the spin-up channel and a pseudogap in the spin-down channel. This behavior is retained qualitatively even after Fe adsorption with a reduction of gap in the spin-up channel. This case is particularly interesting because a half-metallic solution with an integer total moment appears. Therefore, one may expect a spin-filtering of the conduction electrons.

Fig. \ref{fig:fe-ml-dos-1} shows comparison of $m_l$ decomposed local DOS (LDOS) on Fe atom for $n=1$ case within PBE and PBE+U formalism. It is interesting to note that within PBE formalism Fe$_1$ on V$_1$ is non magnetic however it is magnetic within PBE+U formalism. To understand the reason, we have analyzed the LDOS of Fe for both PBE and PBE+U formalism (seen in fig. \ref{fig:fe-ml-dos-1}). The analysis of $m_l$ decomposed LDOS on Fe atom shows that the contribution from the components except $d$ is negligible. Also, the characteristics of the electronic structure are very different from that of pure graphene. The presence of midgap states of $p$ orbital character in presence of a divacancy was discussed before \cite{coleman, jafri}. Here, these states are mainly of $d$ character. So, the characteristics of transport properties are expected to be different. As seen from the figure, the contribution from the $d$ orbital is concentrated in the energy interval of about 0.6 eV below from the Fermi energy. More interestingly, the contribution from spin-up and spin-down channels of Fe atom is substantially different. The contribution near the Fermi level arising from the orbital with $d_{z^{2}-r^{2}}$ character increases, maintaining the difference in the spin channels as the cluster size increases. Such features are of particular interest in spintronic applications with connection to the spin-dependent transport properties.  Thus it is clear that Fe atoms induce a substantial number of states on the Fermi level and it is seen that those are mainly $d$-like. This feature is general and is present in all the clusters.  The state represents a complex of $d$ states on the central Fe atom and heavily distorted $p$ orbitals on neighboring C atoms. It is the interaction between the four C atoms with the Fe cluster which is responsible for the strong binding. 

\subsubsection{\label{sub2sec:spin-dipole}Spin-dipole moments }
\begin{table}[ht]
\caption{$\rm m_s$, $\rm 7<T_z>$ and $\rm m_{eff} = m_s + 7<T_z>$ values for Fe$_n$ clusters adsorbed in n-vacancy graphene system. Calculation are done within PBE+U formalism.\\~\\}
\label{tab-sdp}
\begin{tabular}{c|cccc}
\hline
\hline
\multicolumn{5}{l}{Cluster size=6}\\
\hline
\hline
Cluster & Atom & $\rm m_s$ & $\rm 7<T_z>$  & $\rm m_{eff}$ \\ 
\hline
\multirow{6}{*}{\includegraphics[scale=0.32]{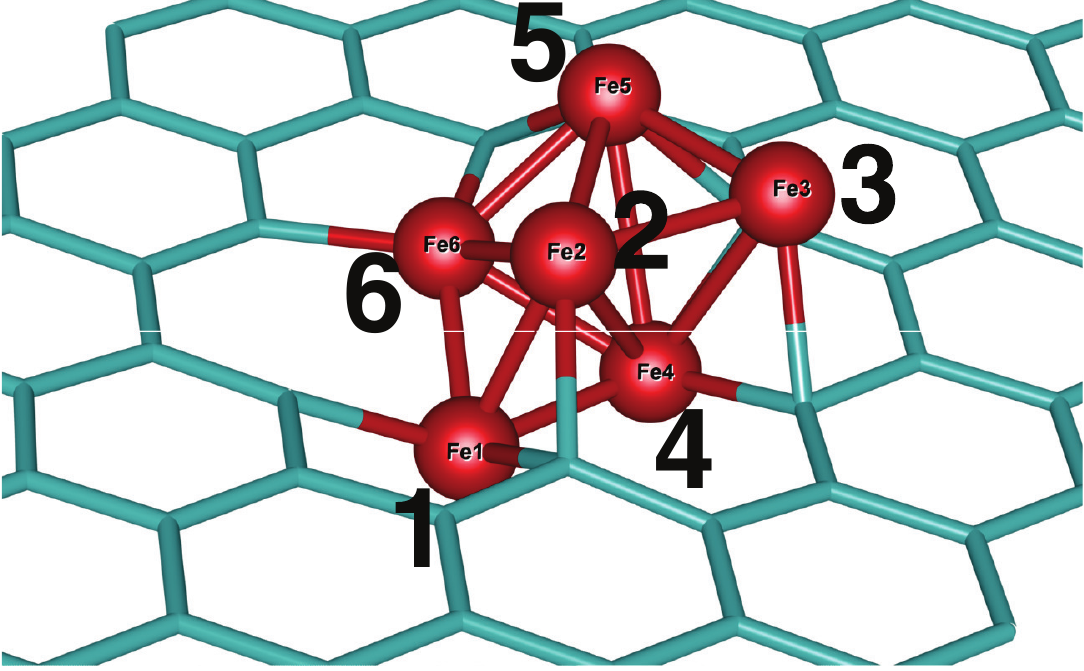}} & 1 & 3.13 & -0.51 & 2.62 \\ 
\cline{2-5}
& 2 & 3.18 & 0.62 & 3.80 \\  \cline{2-5}
& 3 & 3.24 & -0.48 & 2.76 \\  \cline{2-5}
& 4 & 2.99 & 0.11 & 3.10 \\  \cline{2-5}
& 5 & 3.21 & 0.14 & 3.35 \\ \cline{2-5}
& 6 & 2.95 & 0.50 & 3.45 \\
\hline
\hline
\multicolumn{5}{l}{Cluster size=5}\\
\hline
\hline
\multirow{6}{*}{\includegraphics[scale=0.32]{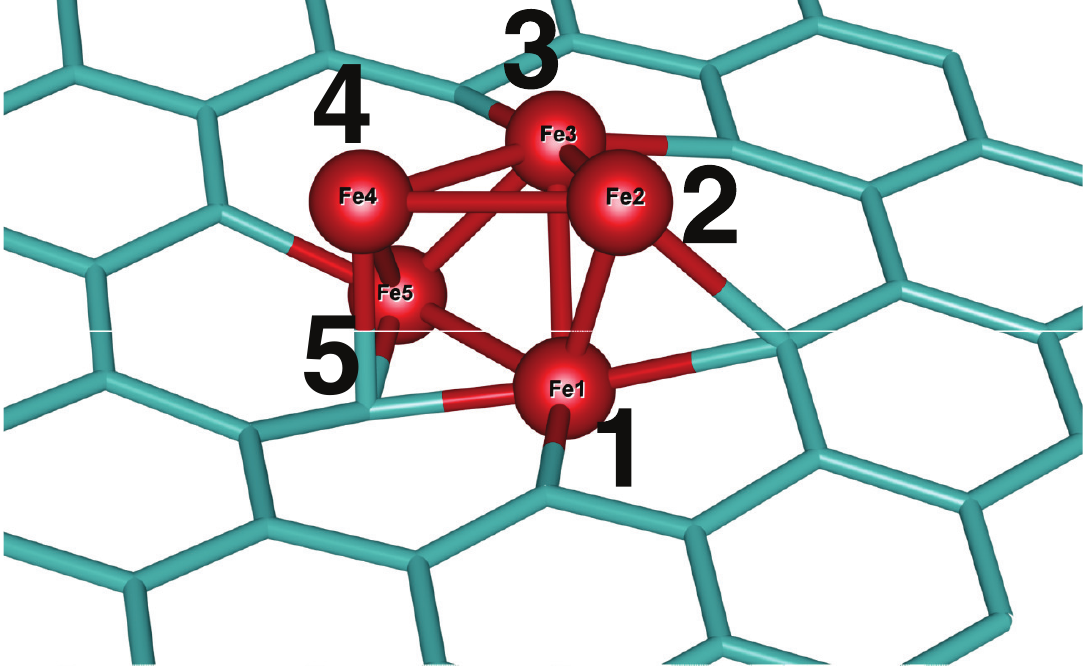}} & 1 & 2.55 & -0.52 & 2.03 \\
\cline{2-5}
& 2 & 3.23 & -0.61 & 2.62 \\
\cline{2-5}
& 3 & 3.06 & 0.65 & 3.71 \\
\cline{2-5}
& 4 & 3.30 & -0.72 & 2.58 \\
\cline{2-5} 
& 5 & 2.73 & -0.41 & 2.32 \\
& & &\\
\hline
\hline
\multicolumn{5}{l}{Cluster size=4}\\
\hline
\hline
\multirow{6}{*}{\includegraphics[scale=0.32]{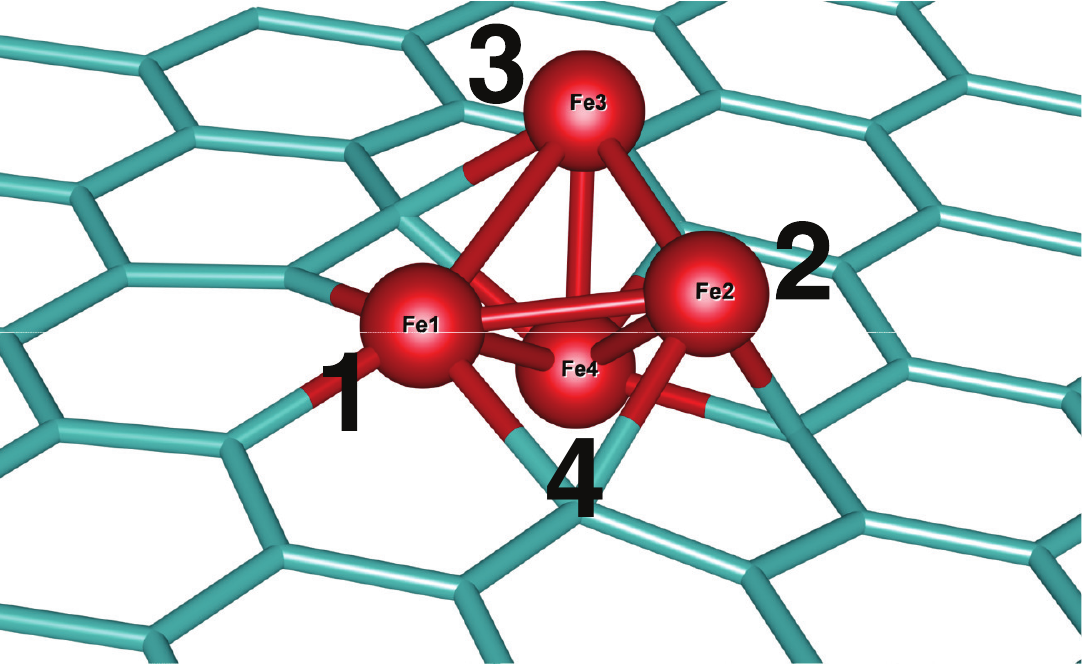}} & 1 & 2.59 & 0.31 & 2.90 \\
\cline{2-5}
& 2 & 3.05 & -0.99 & 2.06 \\
\cline{2-5}
& 3 & 3.14 & -0.44 & 2.70 \\
\cline{2-5}
& 4 & 2.81 & 0.04 & 2.85 \\
& & &\\
& & &\\
\hline 
\hline
\multicolumn{5}{l}{Cluster size=3}\\
\hline
\hline
\multirow{6}{*}{\includegraphics[scale=0.32]{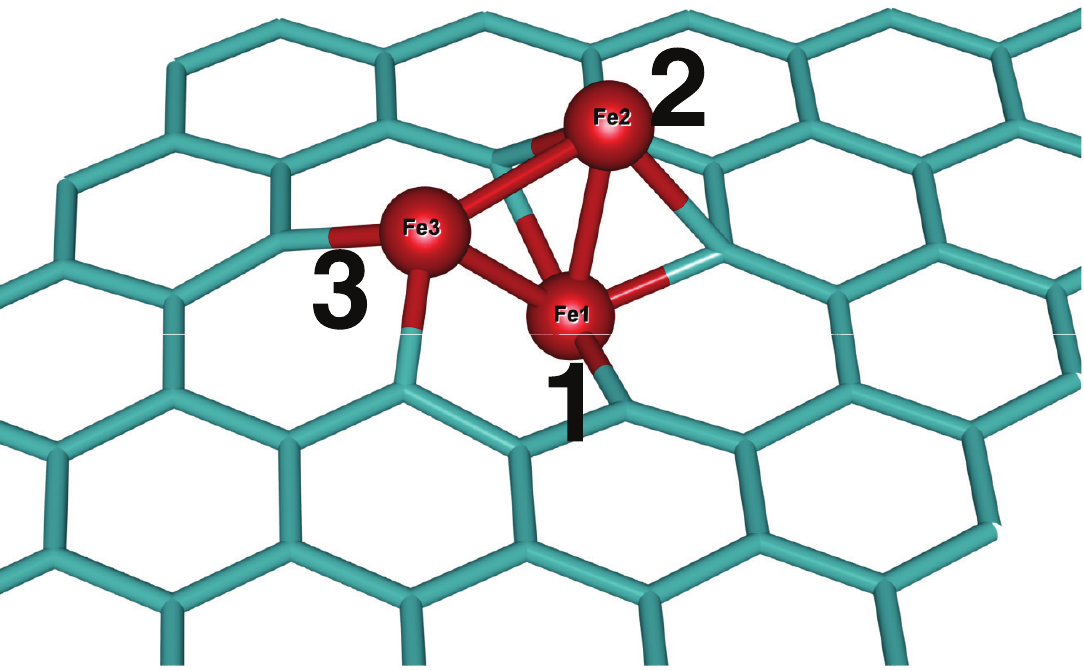}} & 1 & 2.81 & -0.03 & 2.78 \\
& & &\\
\cline{2-5}
& 2 & 3.20 & -0.68 & 2.52 \\
& & &\\
\cline{2-5}
& 3 & 2.85 & 0.31 & 3.16 \\
& & &\\
\hline
\hline
\multicolumn{5}{l}{Cluster size=2}\\
\hline
\hline
\multirow{6}{*}{\includegraphics[scale=0.32]{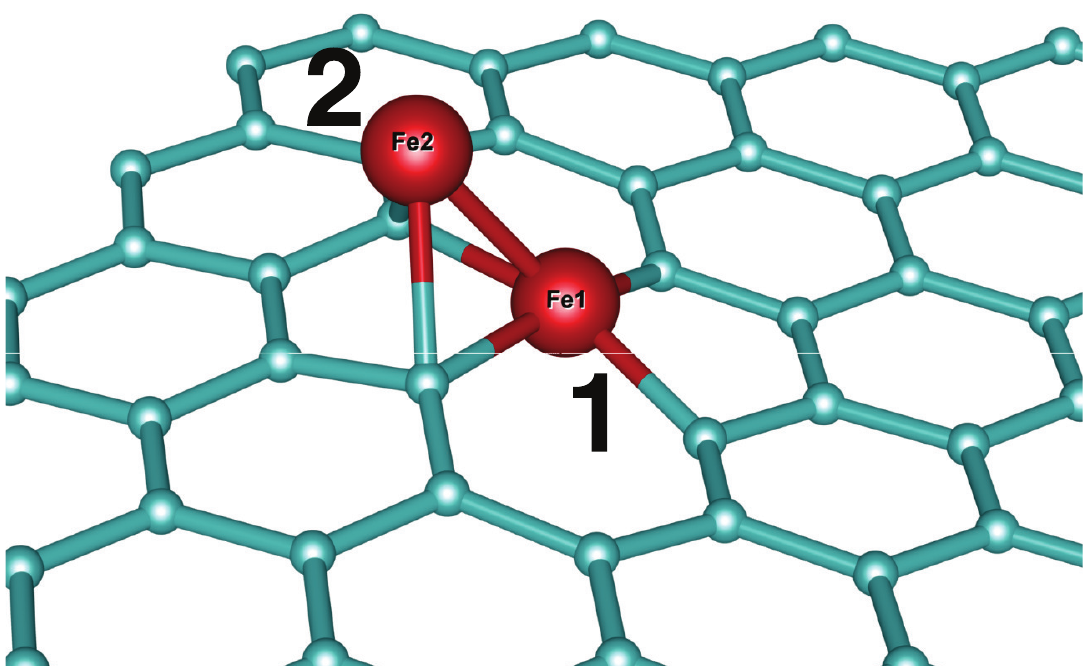}} & 1 & 2.84 & 0.32 & 3.16 \\
& & &\\
& & &\\
\cline{2-5}
& 2 & 2.93 & 0.14 & 3.07 \\
& & &\\
& & &\\
\hline
\hline
\multicolumn{5}{l}{Cluster size=1}\\
\hline
\hline
-- & 1 & 0.91 & -1.15 & -0.24 \\
\hline
\hline
\end{tabular}
\end{table}
So far, we have only discussed the spin moments of Fe. The trapped Fe-clusters are different in sizes and also the environments are different. At the same time, Fe atoms in a particular cluster also experience different crystal fields. Crystal fields from lighter atoms, like C, mostly follow point charge model whereas in the vicinity of Fe, a high bonding effect is expected. So the resultant crystal field can be quite different from each other. In systems having low symmetry, the contribution of the spin-dipole term is expected to be quite significant as discussed in the introduction. The calculated values of $T_z$ depict the behavior of the crystal fields as discussed above. 

The spin dipole operator can be defined as (Oguchi \etal) \cite{oguchi}
 \begin{equation}
    T=\sum_{i}Q^{(i)}s^{(i)},
  \end{equation}
  where, $Q^{(i)}$ is the quadrupolar tensor and can be described as :   
    \begin{eqnarray}
    Q^{(i)}_{\alpha \beta}=\delta_{\alpha \beta}-3\hat{r}^{(i)}_{\alpha}\hat{r}^{(i)}_{\beta}
    \end{eqnarray}    
  Every component of $T$ can be written in second quantization form as   
 \begin{eqnarray}
         T_{\pm}=T_x\pm iT_y=\sum_{\gamma \gamma^{'}}T^{\pm}_{\gamma \gamma^{'}}a^{\dagger}_{\gamma}a_{\gamma^{'}}
          T_z=\sum_{\gamma \gamma^{'}}T^{z}_{\gamma \gamma^{'}}a^{\dagger}_{\gamma}a_{\gamma^{'}}
  \end{eqnarray}  
The matrix elements of $T_{\pm}$ and $T_z$ are :
   \begin{eqnarray}
           T^{\pm}_{\gamma \gamma^{'}}=\bra{\gamma}c^{2}_0s_{\pm}-\sqrt{6}c^{2}_{\pm2}s_{\mp}\pm \sqrt{6}c^{2}_{\pm1}s_{z}\ket{\gamma^{'}}\\
            T^{z}_{\gamma \gamma^{'}}=\bra{\gamma}-\sqrt{\frac {3}{2}} c^{2}_{-1}s_{+}+\sqrt{\frac {3}{2}}c^{2}_{1}s_{-}-2c^{2}_0s_{z}\ket{\gamma^{'}}
    \end{eqnarray}
$\ket{\gamma}=\ket{lm,\sigma}$. 

We have performed DFT calculations, which include all these effects, to calculate spin dipole moments, following the method prescribed by Freeman {\it et. al.} \cite{freeman}. 
In Table \ref{tab-sdp}, the values of spin moments $m_{s}$, spin-dipole moments ($7<T_z>$) and effective moments $m_{eff}$) are shown. One can clearly see that (i) the spin-dipole contributions are not negligible and (ii) the signs of $T_z$ are opposite to $m_s$ in many cases, thereby reducing the effective moments of certain Fe atoms in each cluster. The average effective moment varies non monotonically e.g. the $<m_{eff}>$ are 2.63 $\mu_B$,  2.65 $\mu_B$ and 3.18 $\mu_B$ for $n=4-6$ respectively. In fact,  the value of $7<T_z>$ can reach up to $33\%$ of the spin moment. However, the average effective magnetic moment of the total system does not vary much from the average spin moment due to mutual cancellation of atomic $7<T_z>$.   

\subsubsection{\label{sub2sec:mae}Orbital moments and magnetic anisotropy energies}
As discussed in the introduction, we have included SOC in the Hamiltonian to calculate the orbital moments and magnetic anisotropy energies (MAEs) for Fe$_n$ cluster adsorbed systems. Here, MAE corresponds to only magneto-crystalline energy originating from spin-orbit coupling while the contribution of shape anisotropy is neglected. 
\begin{table}[hbt]
\caption{MAE-Data with PBE+U , magnetic anisotropy
energy ($\Delta E = E^{hard}-E^{easy}$), average orbital moment for Fe$_n$ clusters on n-vacancy graphene}
\label{tab-mae-u}
\centering
\begin{tabular}{|c|c|c|c|c|}
\hline
 Cluster & $<\mu_{orb}>$ & $\Delta E$ & Easy & Hard \\
              &       &  (meV) & axis & axis \\
 \hline
 Fe$_1$ & 0.010 & 0.012 & (001) & (100) \\
 Fe$_2$ & 0.047 & 0.396 & (010) & (100) \\
 Fe$_3$ & 0.042 & 0.843 & (010) & (100) \\
 Fe$_4$ & 0.032 & 0.504 & (001) & (100) \\
 Fe$_5$ & 0.042 & 0.694 & (100) & (010) \\
 Fe$_6$ & 0.042 & 0.708 & (100) & (001) \\
 \hline
\end{tabular}
\end{table}
Table~\ref{tab-mae-u} shows orbital magnetic moments along the easy axes, magnetic anisotropy energies and easy axes. The  calculated average orbital moment of the Fe atoms in these systems are similar to the calculated value for bulk Fe ($\sim$ 0.05 $\mu_B$ without orbital polarization) in the bcc phase. It is interesting to note that Fe$_{1}$ and Fe$_{4}$ clusters exhibit out-of-plane easy axes whereas the other ones have easy in-plane magnetizations. Therefore, for a flux of adatoms of Fe deposited on graphene defect sites with varieties of cluster formation, one may expect the cluster macrospins to lie in the plane or perpendicular to the plane of graphene. However, one should note that the cluster magnetization directions will be quite robust as the magnetic anisotropy energies are not so small in magnitude.  

\section{\label{sec:conclude}Conclusions}

We have performed systematic studies of the evolution of structural, electronic and magnetic properties of Fe$_n$ ($n=1-6$) clusters on graphene sheets having different types of vacancies using DFT combined with Coulomb correlations for Fe-d electrons within the Hubbard formalism. It has been found that the vacancy formation energy decreases as a function of the size of the vacancy hole and hence, the formation of correlated vacancy is facilitated. Our ab initio molecular dynamics simulations suggest that the adsorbed Fe adatoms form clusters in a very short time scale and the clusters get stuck at the vacancy sites. Moreover, it is observed that Fe clusters promote vacancy formation, i.e., it comes easier to remove C atoms from the graphene lattice in presence of Fe. The strong adsorption of Fe clusters at the vacancy sites gives rise to anisotropy in geometries and spin densities. Hence, a strong variation of local spin-dipole moments (parallel or antiparallel to the spin moments) yields a significant variation in the effective moments, which are measurable in XMCD experiments. Finally, our calculations of magnetic anisotropy energies reveal that for some of the cluster sizes, out-of-plane easy axes of magnetization are stabilized with a large MAE, which could be important for magnetic data storage.

\section*{Acknowledgment} 
S.H., B.S. and O.E. would like to acknowledge KAW foundation for financial support. In
addition, B.S. acknowledges Carl Tryggers Stiftelse and Swedish Research Council
for financial support. O.E. acknowledges support from VR, eSSENCE and the ERC (project 247062 - ASD). We thank SNIC-NSC and SNIC-HPC2N computing centers under Swedish National Infrastructure for Computing (SNIC) and the PRACE-2IP project (FP7 RI-283493)
resource Abel supercomputer based in Norway at University of Oslo for granting computer time.

 \bibliography{biblio}

\end{document}